\documentclass[preprintnumbers,prd,twocolumn,showpacs,floatfix,preprintnumbers,letterpaper,amsmath,amssymb,nofootinbib,superscriptaddress]{revtex4}
\usepackage{graphicx}
\usepackage{epsfig}
\usepackage{bm}
\usepackage{amsfonts}

\newcommand{\la}{\lambda}

\newcommand{\bear}{\begin{eqnarray}}

\newcommand{\eear}{\end{eqnarray}}

\newbox\pippobox

\def\6{\partial}

\def\a{\alpha}

\def\sq
\def\a{\alpha}

\def\bx{{\bf x}}

\begin{document}

\title{Phase-space analysis of Ho\v{r}ava-Lifshitz cosmology}

\author{Genly Leon}
\email{genly@uclv.edu.cu} \affiliation{Department of Mathematics,
Universidad Central de Las Villas, Santa Clara CP 54830, Cuba}

\author{Emmanuel N. Saridakis }
\email{msaridak@phys.uoa.gr} \affiliation{Department of Physics,
University of Athens, GR-15771 Athens, Greece}

\begin{abstract}
We perform a detailed phase-space analysis of Ho\v{r}ava-Lifshitz
cosmology, with and without the detailed-balance condition. Under
detailed-balance we find that the universe can reach a
bouncing-oscillatory state at late times, in which dark-energy,
behaving as a simple cosmological constant, is dominant. In the
case where the detailed-balance condition is relaxed, we find that
the universe reaches an eternally expanding, dark-energy-dominated
solution, with the oscillatory state preserving also a small
probability. Although this analysis indicates that
Ho\v{r}ava-Lifshitz cosmology can be compatible with observations,
it does not enlighten the discussion about its possible conceptual
and theoretical problems.
\end{abstract}

 \pacs{98.80.Cq, 04.50.Kd}

\maketitle

\section{Introduction}

Recently, a power-counting renormalizable, ultra-violet (UV)
complete theory of gravity was proposed by Ho\v{r}ava in
\cite{hor2,hor1,hor3,hor4}. Although presenting an infrared (IR)
fixed point, namely General Relativity, in the  UV the theory
possesses a fixed point with an anisotropic, Lifshitz scaling
between time and space of the form $\bx\to\ell~\bx$,
$t\to\ell^z~t$, where $\ell$, $z$, $\bx$ and $t$ are the scaling
factor, dynamical critical exponent, spatial coordinates and
temporal coordinate, respectively.

Due to these novel features, there has been a large amount of
effort in examining and extending the properties of the theory
itself
\cite{Volovik:2009av,Cai:2009ar,Cai:2009dx,Orlando:2009en,Nishioka:2009iq,Konoplya:2009ig,Charmousis:2009tc,Li:2009bg,Visser:2009fg,Sotiriou:2009bx,Chen:2009bu,Chen:2009ka,Shu:2009gc,Bogdanos:2009uj,Kluson:2009rk,Afshordi:2009tt,Myung:2009ur}.
Additionally, application of Ho\v{r}ava-Lifshitz gravity as a
cosmological framework gives rise to Ho\v{r}ava-Lifshitz
cosmology, which proves to lead to interesting behavior
\cite{Calcagni:2009ar,Kiritsis:2009sh}. In particular, one can
examine specific solution subclasses
\cite{Lu:2009em,Nastase:2009nk,Colgain:2009fe,Ghodsi:2009rv,Minamitsuji:2009ii,Ghodsi:2009zi},
the perturbation spectrum
\cite{Mukohyama:2009gg,Piao:2009ax,Gao:2009bx,Chen:2009jr,Gao:2009ht,Wang:2009yz,Kobayashi:2009hh},
the gravitational wave production
\cite{Mukohyama:2009zs,Takahashi:2009wc,Koh:2009cy}, the matter
bounce
\cite{Brandenberger:2009yt,Brandenberger:2009ic,Cai:2009in}, the
black hole properties
\cite{Danielsson:2009gi,Cai:2009pe,Myung:2009dc,Kehagias:2009is,Cai:2009qs,Mann:2009yx,Bertoldi:2009vn,Castillo:2009ci,BottaCantcheff:2009mp,Lee:2009rm},
the dark energy phenomenology
\cite{Saridakis:2009bv,Park:2009zr,Wang:2009rw,Appignani:2009dy},
the astrophysical phenomenology \cite{Kim:2009dq,Harko:2009qr}
etc. However, despite this extended research, there are still many
ambiguities if Ho\v{r}ava-Lifshitz gravity is reliable and capable
of a successful description of the gravitational background of our
world, as well as of the cosmological behavior of the universe
\cite{Charmousis:2009tc,Sotiriou:2009bx,Bogdanos:2009uj}.

Although the discussion about the foundations and the possible
conceptual and phenomenological problems of Ho\v{r}ava-Lifshitz
gravity and cosmology is still open in the literature, it is worth
investigating in a systematic way the possible cosmological
behavior of a universe governed by Ho\v{r}ava gravity. Thus, in
the present work we perform a phase-space and stability analysis
of Ho\v{r}ava-Lifshitz cosmology, with or without the
detailed-balance condition, and we are interesting in
investigating the possible late-time solutions. In these solutions
we calculate various observable quantities, such are the
dark-energy density and equation-of-state parameters. As we see,
indeed Ho\v{r}ava-Lifshitz cosmology can be consistent with
observations and in addition it can give rise to a bouncing
universe. Furthermore, the results seem to be independent of the
specific form of the dark matter content of the universe. This
analysis however does not enlighten the discussion about possible
conceptual problems and instabilities of Ho\v{r}ava-Lifshitz
gravity, which is the subject of interest of other studies.

The paper is organized as follows: In section \ref{model} we
present the basic ingredients of Ho\v{r}ava-Lifshitz cosmology,
extracting the Friedmann equations, and describing the dark matter
and dark energy dynamics. In section \ref{phaseanalysis} we
perform a systematic phase-space and stability analysis for
various cases under the detailed-balance condition, including a
flat or non-flat geometry in the presence or not of a cosmological
constant. In section \ref{nondetbal} we extend the phase-space
analysis in the case where the detailed-balance condition is
relaxed. In section \ref{cosmimpl} we discuss the corresponding
cosmological implications and the effects on observable
quantities. Finally, our results are summarized in section
\ref{conclusions}.

\section{Ho\v{r}ava-Lifshitz cosmology}
\label{model}

 We begin with a brief review of Ho\v{r}ava-Lifshitz
gravity. The dynamical variables are the lapse and shift
functions, $N$ and $N_i$ respectively, and the spatial metric
$g_{ij}$ (roman letters indicate spatial indices). In terms of
these fields the full metric is
\begin{eqnarray}
ds^2 = - N^2 dt^2 + g_{ij} (dx^i + N^i dt ) ( dx^j + N^j dt ) ,
\end{eqnarray} 
where indices are raised and lowered using $g_{ij}$. The scaling
transformation of the coordinates reads (z=3):
\begin{eqnarray}
 t \rightarrow l^3 t~~~{\rm and}\ \ x^i \rightarrow l x^i~.
\end{eqnarray}

Decomposing the gravitational action into a kinetic and a
potential part as $S_g = \int dt d^3x \sqrt{g} N ({\cal L}_K+{\cal
L}_V)$, and under the assumption of detailed balance \cite{hor3}
(the extension beyond detail balance will be performed later on),
which apart form reducing the possible terms in the Lagrangian it
allows for a quantum inheritance principle \cite{hor2} (the $D+1$
dimensional theory acquires the renormalization properties of the
D-dimensional one),
 the full action of Ho\v{r}ava-Lifshitz gravity is given by
\begin{eqnarray}
S_g =  \int dt d^3x \sqrt{g} N \left\{ \frac{2}{\kappa^2}
(K_{ij}K^{ij} - \lambda K^2)- \ \ \ \ \ \ \ \ \ \ \ \ \ \ \ \ \  \right. \nonumber \\
\left.
 - \frac{\kappa^2}{2 w^4} C_{ij}C^{ij}
 + \frac{\kappa^2 \mu}{2 w^2}
\frac{\epsilon^{ijk}}{\sqrt{g}} R_{il} \nabla_j R^l_k -
\frac{\kappa^2 \mu^2}{8} R_{ij} R^{ij}+
     \right. \nonumber \\
\left.    + \frac{\kappa^2 \mu^2}{8(1 - 3 \lambda)} \left[ \frac{1
- 4 \lambda}{4} R^2 + \Lambda  R - 3 \Lambda ^2 \right] \right\},
\end{eqnarray}
where
\begin{eqnarray}
K_{ij} = \frac{1}{2N} \left( {\dot{g_{ij}}} - \nabla_i N_j -
\nabla_j N_i \right) \, ,
\end{eqnarray}
is the extrinsic curvature and
\begin{eqnarray} C^{ij} \, = \, \frac{\epsilon^{ijk}}{\sqrt{g}} \nabla_k
\bigl( R^j_i - \frac{1}{4} R \delta^j_i \bigr)
\end{eqnarray}
the Cotton tensor, and the covariant derivatives are defined with
respect to the spatial metric $g_{ij}$. $\epsilon^{ijk}$ is the
totally antisymmetric unit tensor, $\lambda$ is a dimensionless
constant and $\Lambda $ is a negative constant which is related to
the cosmological constant in the IR limit. Finally, the variables
$\kappa$, $w$ and $\mu$ are constants with mass dimensions $-1$,
$0$ and $1$, respectively.

In order to add the dark-matter content in a universe governed by
Ho\v{r}ava gravity, a scalar field is introduced
\cite{Calcagni:2009ar,Kiritsis:2009sh}, with action:
\begin{eqnarray}
S_M\equiv S_\phi = \int dtd^3x \sqrt{g} N \left[
\frac{3\lambda-1}{4}\frac{\dot\phi^2}{N^2}
+m_1m_2\phi\nabla^2\phi-\right.\nonumber\\
\left.-\frac{1}{2}m_2^2\phi\nabla^4\phi +
\frac{1}{2}m_3^2\phi\nabla^6\phi -V(\phi)\right],\ \ \ \ \
\end{eqnarray}
where $V(\phi)$ acts as a potential term and $m_i$ are constants.
Although one could just follow a hydrodynamical approximation and
introduce straightaway the density and pressure of a matter fluid
\cite{Sotiriou:2009bx}, the field approach is more robust,
especially if one desires to perform a phase-space analysis.

Now, in order to focus on cosmological frameworks, we impose the
so called projectability condition \cite{Charmousis:2009tc} and
use an FRW metric,
\begin{eqnarray}
N=1~,~~g_{ij}=a^2(t)\gamma_{ij}~,~~N^i=0~,
\end{eqnarray}
with
\begin{eqnarray}
\gamma_{ij}dx^idx^j=\frac{dr^2}{1-kr^2}+r^2d\Omega_2^2~,
\end{eqnarray}
where $k=-1,0,1$ correspond to open, flat, and closed universe
respectively. In addition, we assume that the scalar field is
homogenous, i.e $\phi\equiv\phi(t)$. By varying $N$ and $g_{ij}$,
we obtain the equations of motion:
\begin{eqnarray}\label{Fr1}
H^2 &=&
\frac{\kappa^2}{6(3\la-1)}\left[\frac{3\la-1}{4}\,\dot\phi^2
+V(\phi)\right]+\nonumber\\
&+&\frac{\kappa^2}{6(3\la-1)}\left[
-\frac{3\kappa^2\mu^2k^2}{8(3\lambda-1)a^4}
-\frac{3\kappa^2\mu^2\Lambda ^2}{8(3\lambda-1)}
 \right]+\nonumber\\
 &+&\frac{\kappa^4\mu^2\Lambda k}{8(3\lambda-1)^2a^2} \ ,
\end{eqnarray}
\begin{eqnarray}\label{Fr2}
\dot{H}+\frac{3}{2}H^2 &=&
-\frac{\kappa^2}{4(3\la-1)}\left[\frac{3\la-1}{4}\,\dot\phi^2
-V(\phi)\right]-\nonumber\\
&-&\frac{\kappa^2}{4(3\la-1)}\left[-\frac{\kappa^2\mu^2k^2}{8(3\lambda-1)a^4}
+\frac{3\kappa^2\mu^2\Lambda ^2}{8(3\lambda-1)}
 \right]+\nonumber\\
 &+&\frac{\kappa^4\mu^2\Lambda k}{16(3\lambda-1)^2a^2}\ ,
\end{eqnarray}
where we have defined the Hubble parameter as $H\equiv\frac{\dot
a}{a}$. Finally, the equation of motion for the scalar field
reads:
\begin{eqnarray}\label{phidott}
&&\ddot\phi+3H\dot\phi+\frac{2}{3\lambda-1}\frac{dV(\phi)}{d\phi}=0.
\end{eqnarray}

At this stage we can define the energy density and pressure for
the scalar field responsible for the matter content of the
Ho\v{r}ava-Lifshitz universe:
\begin{eqnarray}
&&\rho_M\equiv \rho_\phi=\frac{3\la-1}{4}\,\dot\phi^2
+V(\phi)\label{rhom}\\
&&p_M\equiv p_\phi=\frac{3\la-1}{4}\,\dot\phi^2
-V(\phi).\label{pressurem}
\end{eqnarray}
Concerning the dark-energy sector we can define
\begin{equation}\label{rhoDE}
\rho_{DE}\equiv -\frac{3\kappa^2\mu^2k^2}{8(3\lambda-1)a^4}
-\frac{3\kappa^2\mu^2\Lambda ^2}{8(3\lambda-1)}
\end{equation}
\begin{equation}
\label{pDE} p_{DE}\equiv
-\frac{\kappa^2\mu^2k^2}{8(3\lambda-1)a^4}
+\frac{3\kappa^2\mu^2\Lambda ^2}{8(3\lambda-1)}.
\end{equation}
The term proportional to $a^{-4}$ is the usual ``dark radiation
term'', present in Ho\v{r}ava-Lifshitz cosmology
\cite{Calcagni:2009ar,Kiritsis:2009sh}. Finally, the constant term
is just the explicit (negative) cosmological constant. Therefore,
in expressions (\ref{rhoDE}),(\ref{pDE}) we have defined the
energy density and pressure for the effective dark energy, which
incorporates the aforementioned contributions.

Using the above definitions, we can re-write the Friedmann
equations (\ref{Fr1}),(\ref{Fr2}) in the standard form:
\begin{eqnarray}
\label{Fr1b} H^2 &=&
\frac{\kappa^2}{6(3\la-1)}\Big[\rho_M+\rho_{DE}\Big]+ \frac{\beta k}{a^2}\\
\label{Fr2b} \dot{H}+\frac{3}{2}H^2 &=&
-\frac{\kappa^2}{4(3\la-1)}\Big[p_M+p_{DE}
 \Big]+ \frac{\beta k}{2a^2}.
\end{eqnarray}
In these relations we have defined
$\beta\equiv\frac{\kappa^4\mu^2\Lambda }{8(3\lambda-1)^2}$, which
is the coefficient of the curvature term. Additionally, we could
also define an effective Newton's constant and an effective light
speed \cite{Calcagni:2009ar,Kiritsis:2009sh}, but we prefer to
keep $\frac{\kappa^2}{6(3\la-1)}$ in the expressions, just to make
clear the origin of these terms in Ho\v{r}ava-Lifshitz cosmology.
Finally, note that using (\ref{phidott}) it is straightforward to
see that the aforementioned dark matter and dark energy quantities
verify the standard evolution equations:
\begin{eqnarray}\label{phidot2}
&&\dot{\rho}_M+3H(\rho_M+p_M)=0\\
\label{sdot2} &&\dot{\rho}_{DE}+3H(\rho_{DE}+p_{DE})=0.
\end{eqnarray}

The aforementioned formulation of Ho\v{r}ava-Lifshitz cosmology
has been performed under the imposition of the detailed-balance
condition. However, in the literature there is a discussion
whether this condition leads to reliable results or if it is able
to reveal the full information of Ho\v{r}ava-Lifshitz
 gravity \cite{Calcagni:2009ar,Kiritsis:2009sh}. Thus, for
 completeness, we add here the Friedmann equation in the case
 where detailed balance is relaxed. In such a case one can in
 general write
 \cite{Charmousis:2009tc,Sotiriou:2009bx,Bogdanos:2009uj}:
\begin{eqnarray}\label{Fr1c}
H^2 &=&
\frac{2\sigma_0}{(3\la-1)}\left[\frac{3\la-1}{4}\,\dot\phi^2
+V(\phi)\right]+\nonumber\\
&+&\frac{2}{(3\la-1)}\left[
\frac{\sigma_1}{6}+\frac{\sigma_3k^2}{6a^4} +\frac{\sigma_4k}{a^6}
 \right]+\nonumber\\&+&\frac{\sigma_2}{3(3\la-1)}\frac{k}{a^2}
\end{eqnarray}
\begin{eqnarray}\label{Fr2c}
\dot{H}+\frac{3}{2}H^2 &=&
-\frac{3\sigma_0}{(3\la-1)}\left[\frac{3\la-1}{4}\,\dot\phi^2
-V(\phi)\right]-\nonumber\\
&-&\frac{3}{(3\la-1)}\left[
-\frac{\sigma_1}{6}+\frac{\sigma_3k^2}{18a^4}
+\frac{\sigma_4k}{6a^6}
 \right]+\nonumber\\&+&
 \frac{\sigma_2}{6(3\la-1)}\frac{k}{a^2} ,
\end{eqnarray}
where $\sigma_0\equiv \kappa^2/12$, and the constants $\sigma_i$
are arbitrary (although one can set $\sigma_2$ to be positive
too). Thus, the effect of the detailed-balance relaxation is the
decoupling of the coefficients, together with the appearance of a
term proportional to $a^{-6}$. This term has a negligible impact
at large scale factors, however it could play a significant role
at small ones. Finally, in the non-detailed-balanced case, the
energy density and pressure for matter coincide with those of
detailed-balance scenario (expressions
(\ref{rhom}),(\ref{pressurem})),  since the detailed-balance
condition affects only the gravitational sector of the theory and
has nothing to do with the matter content of the universe.
However, the corresponding quantities for dark energy are
generalized to
\begin{eqnarray}\label{rhoDEext}
&&\rho_{DE}|_{_\text{non-db}}\equiv
\frac{\sigma_1}{6}+\frac{\sigma_3k^2}{6a^4} +\frac{\sigma_4k}{a^6}
\\
&&\label{pDEext} p_{DE}|_{_\text{non-db}}\equiv
-\frac{\sigma_1}{6}+\frac{\sigma_3k^2}{18a^4}
+\frac{\sigma_4k}{6a^6}.
\end{eqnarray}

Having presented the cosmological equations of a universe governed
by Ho\v{r}ava-Lifshitz gravity, under detailed-balance condition
or not, we can investigate the possible cosmological behaviors and
discuss the corresponding physical implications by performing a
phase-space analysis. This is done in the following two sections,
for the detailed and non-detailed balance cases separately.

\section{Detailed balance: Phase-space analysis}
\label{phaseanalysis}

In order to perform the phase-space and stability analysis of the
Ho\v{r}ava-Lifshitz universe, we have to transform the
cosmological equations into an autonomous dynamical system
\cite{Copeland:1997et,expon,expon1,expon2}. This will be achieved
by introducing the auxiliary variables:
\begin{eqnarray}
&&x=\frac{\kappa  \dot \phi }{2 \sqrt{6} H}, \label{auxilliaryx}\\
&&y=\frac{\kappa \sqrt{V(\phi)}}{\sqrt{6}H \sqrt{3 \lambda -1}} \label{auxilliaryy}\\
&&z=\frac{\kappa ^2 \mu }{4 (3 \lambda -1) a^2 H} \label{auxilliaryz}\\
&&u=\frac{\kappa ^2 \Lambda  \mu }{4 (3 \lambda -1) H},
 \label{auxilliaryu}
\end{eqnarray}
together with $M=\ln a$. Thus, it is easy to see that for every
quantity $F$ we acquire $\dot{F}=H\frac{dF}{dM}$.
 Using these
variables we can straightforwardly obtain the density parameters
of dark matter and dark energy (through expressions (\ref{rhom}),
(\ref{rhoDE})) as:
\begin{eqnarray}
&&
\Omega_M\equiv\frac{\kappa^{2}}{6(3\lambda-1)H^{2}}\rho_M=x^2+y^2,
 \label{OmegaM}\\
&&
\Omega_{DE}\equiv\frac{\kappa^{2}}{6(3\lambda-1)H^{2}}\rho_{DE}=-k^2z^2-u^2
 \label{OmegaDE},
\end{eqnarray}
 and in addition we can calculate the corresponding
 equation-of-state parameters:
\begin{eqnarray}
&& w_M\equiv\frac{p_M}{\rho_M}=\frac{x^2-y^2}{x^2+y^2},
 \label{wM}\\
 && w_{DE}\equiv\frac{p_{DE}}{\rho_{DE}}=\frac{k^2z^2-3u^2}{3k^2z^2+3u^2}
 \label{wDE}.
\end{eqnarray}
We mention that these relations are always valid, that is
independently of the specific state of the system (they are valid
in the whole phase-space and not only at the critical points).
Finally, for completeness, and observing (\ref{Fr1b}), we can
define the curvature density parameter as:
\begin{eqnarray}
&& \Omega_k\equiv\frac{\beta k}{H^2a^2}=2kuz.
 \label{OmegaK}
\end{eqnarray}

Using the auxiliary variables
(\ref{auxilliaryx}),(\ref{auxilliaryy}),(\ref{auxilliaryz}),(\ref{auxilliaryu})
the cosmological equations of motion (\ref{Fr1b}), (\ref{Fr2b}),
(\ref{phidot2}) and (\ref{sdot2}), can be transformed into an
autonomous form
 $\textbf{X}'=\textbf{f(X)}, $ where $\textbf{X}$ is the column
vector constituted by the auxiliary variables, \textbf{f(X)} the
corresponding  column vector of the autonomous equations, and
prime denotes derivative with respect to $M=\ln a$. The critical
points $\bf{X_c}$ are extracted satisfying $\bf{X}'=0$, and in
order to determine the stability properties of these critical
points we expand around $\bf{X_c}$, setting
$\bf{X}=\bf{X_c}+\bf{U}$ with $\textbf{U}$ the perturbations of
the variables considered as a column vector. Thus, up to the first
order we acquire $ \textbf{U}'={\bf{Q}}\cdot \textbf{U}, $ where
the matrix ${\bf {Q}}$ contains the coefficients of the
perturbation equations. Thus, for each critical point, the
eigenvalues of ${\bf {Q}}$ determine its type and stability.

In the following we perform a phase-space analysis of the
cosmological system at hand. As we can see from the Friedmann
equations (\ref{Fr1}), (\ref{Fr2}) one can have a zero or non-zero
cosmological constant, in a flat or non-flat universe. Thus, for
simplicity we investigate separately the corresponding four cases.
Finally, note that we assume $\la
>\frac{1}{3}$ as required by the consistency of the Ho\v{r}ava
gravitational background, but we do not impose any other
constraint on the model parameters (although one could do so using
the light speed and Newton's constant values) in order to remain
as general as possible.

\subsection{Case 1: Flat universe with $\Lambda=0$}\label{section A}

In this scenario the variable $u$ is irrelevant, and the Friedmann
equations (\ref{Fr1}), (\ref{Fr2}) become:
\begin{eqnarray}
&&1=x^2+y^2 \label{Fr1bcase1}
\\
&&\frac{H'}{H}=-3 x^2. \label{Fr2bcase1}
\end{eqnarray}
Thus, after using the first of these relations in order to
eliminate one variable, the corresponding autonomous system
writes:
\begin{eqnarray}
x'&=&\left(3 x-\sqrt{6} q\right) \left(x^2-1\right),\label{eqxcase1}\nonumber\\
z'&=&  \left(3 x^2-2\right) z.\label{eqzcase1}
\end{eqnarray}
We mention that for simplicity we have set $q=-\frac{1}{\kappa
V(\phi)}\frac{d V(\phi)}{d\phi}$ and we have assumed it to be a
constant, that is we are investigating the usual exponential
potentials. However, as we will see this is not necessary, since
the most important results of the present work are independent of
the matter sector.

The autonomous system (\ref{eqzcase1}) is defined in the phase
space $$\Psi=\left\{(x,z): -1\leq x\leq 1,
z\in\mathbb{R}\right\}.$$ As we observe, this phase plane is not
compact since $z$ is in general unbounded. However, the system is
integrable and the orbit in the plane $\Psi$, passing initially
through $(x_0,\,z_0)$, can be obtained explicitly and it is given
by the graph
\begin{eqnarray}
z(x)=z_0\left(\frac{3 x-\sqrt{6}
   q}{3
   x_0-\sqrt{6} q}\right)^{1+\frac{1}{2 q^2-3}}
   \left(\frac{x^2-1}{x_0^2-1}\right)^{\frac{1}{6-4
   q^2}}\cdot\nonumber\\
  \cdot \exp\left\{\frac{\sqrt{6} q \left[\tanh ^{-1}(x)-\tanh ^{-1}(x_0)\right]}{6
   q^2-9}\right\}.
\end{eqnarray}
The critical points $(x_c,z_c)$ of the autonomous system
(\ref{eqzcase1})  are obtained by setting the left hand sides of
the equations to zero. They are displayed in table \ref{crit},
where we also present the necessary conditions for their
existence. In addition, for each critical point we calculate the
values of $w_{M}$ (given by relation (\ref{wM})), of $\Omega_{DE}$
(given by (\ref{OmegaDE})), and of $w_{DE}$ (given by
(\ref{wDE})). Note that in this case, $w_{DE}$ remains unspecified
and the results hold independently of its value. Finally, The
$2\times2$ matrix ${\bf {Q}}$ of the linearized perturbation
equations writes:
\[{\bf {Q}}= \left[ \begin{array}{cc}
 9 x^2-2 \sqrt{6} q x-3 & 0 \\
 6 x z & 3 x^2-2
\end{array} \right],\]
and in  table \ref{crit} we display its eigensystems (eigenvalues
and associated eigenvectors) evaluated at each critical points,
 as
well as their type and stability, acquired by examining the sign
of the real part of the eigenvalues.
\begin{table*}[ht]
\begin{center}
\begin{tabular}{ccccccccc}
\hline \hline
 Cr. P &$x_c$&$z_c$&\quad Existence\quad& Eigensystem&\quad Stable for&\quad $w_M$&\quad $\Omega_{DE}$&\quad $w_{DE}$\\
\hline $P_{1,2}$& $\pm 1$ &0&\quad All $q$ \quad & \quad $
\begin{array}{cc}
  \mp 2 \sqrt{6} q+6 & 1\\
  \{1,0\} & \{0,1\}
\end{array}
$&\quad unstable&\quad1&\quad0&\quad arbitrary\quad \vspace{0.2cm}\\
\hline $P_{3}$& $\sqrt{\frac{2}{3}}\,q$ &0&\quad
$-\sqrt{\frac{3}{2}}<q<\sqrt{\frac{3}{2}}$ \quad & \quad $
\begin{array}{cc}
  -3+2 q^2 & 2\left(-1+q^2\right)\\
  \{1,0\} & \{0,1\}
\end{array}
$& \quad $-1<q<1$ &\quad$\frac{4}{3}q^2-1$&\quad0&arbitrary\quad \\
\vspace{-0.35cm}\\
\hline $P_{4}$ & $\sqrt{\frac{2}{3}} q$ &$z_c$&\quad $q=\pm 1$
\quad & \quad $
\begin{array}{cc}
 -1 & 0 \\
 \left\{\mp\frac{1}{2 \sqrt{6} z_c},1\right\} & \{0,1\}
\end{array}
$& nonhyperbolic &\quad$1/3$&\quad0&arbitrary\quad \\
\vspace{-0.35cm}\\
\hline \hline
\end{tabular}
\end{center}
\caption[crit]{\label{crit} The critical points of  a flat
universe with $\Lambda=0$ (case 1) and their behavior.}
\end{table*}

 For hyperbolic critical
points (all the eigenvalues have real parts different from zero)
one can easily extract their type (source (unstable) for positive
real parts, saddle for real parts of different sign and sink
(stable) for negative real parts). However, if at least one
eigenvalue has a zero real part (non-hyperbolic critical point)
one is not able to obtain conclusive information about the
stability from linearization and needs to resort to other tools
like Normal Forms calculations  \cite{arrowsmith,wiggins}, or
numerical experimentation.
\begin{figure}[ht]
\begin{center}
\mbox{\epsfig{figure=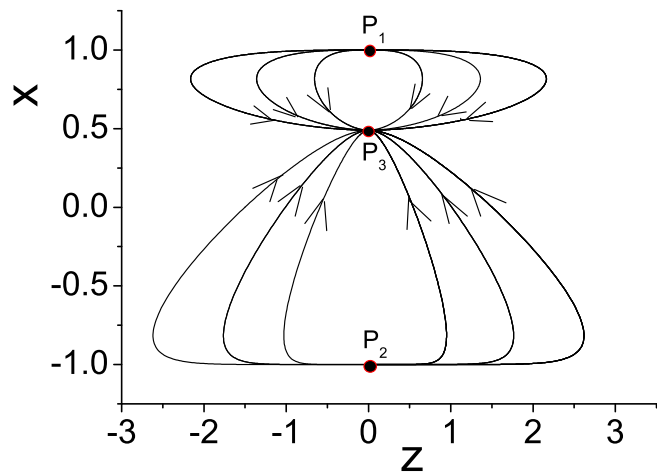,width=8.8cm,angle=0}} \caption{ {\it
Phase plane for a flat universe with $\Lambda=0$ (case 1), for the
choice $q=0.6.$ The critical points $P_{1}$ and $P_2$ are unstable
(sources), while $P_3$ is a global attractor.}} \label{Fig1}
\end{center}
\end{figure}
Thus, in the case at hand, $P_1$ and $P_2$ are nonhyperbolic for
$q= \sqrt{3/2},$ while $P_3$ is nonhyperbolic for $q^2\in \{3/2,
1\}$. Finally, note that in the special case where $q=\pm1$, the
system admits an extra curve of critical points $P_4.$ Each point
in $P_4$ is non-hyperbolic, with center manifold tangent to the
z-axis, but the curve $P_4$ is actually ``normally hyperbolic''
\cite{normally}. This means that we can indeed analyze the
stability by analyzing the sign of the real parts of the non-null
eigenvalues. Therefore, since the non zero eigenvalue is negative,
$P_4$ is a local attractor.
\begin{table*}[t]
\begin{center}
\begin{tabular}{ccccccccc}
\hline \hline
 Cr. P &$x_c$&$z_c$&\quad Existence\quad& Eigensystem&\quad Stable for&\quad $w_M$&\quad $\Omega_{DE}$&\quad $w_{DE}$\\
\hline $P_{1,2}$& $\pm 1$ &0&\quad All $q$ \quad & \quad $
\begin{array}{cc}
  \mp 2 \sqrt{6} q+6 & 1\\
  \{1,0\} & \{0,1\}
\end{array}
$&\quad unstable&\quad1&\quad0&\quad arbitrary\quad \vspace{0.2cm}\\
\hline $P_{3}$& $\sqrt{\frac{2}{3}}\,q$ &0&\quad
$-\sqrt{\frac{3}{2}}<q<\sqrt{\frac{3}{2}}$ \quad & \quad $
\begin{array}{cc}
  -3+2 q^2 & 2\left(-1+q^2\right)\\
  \{1,0\} & \{0,1\}
\end{array}
$& \quad $-1<q<1$ &\quad$\frac{4}{3}q^2-1$&\quad0&arbitrary\quad \\
\vspace{-0.35cm}\\
\hline $P_{5,6}$& $\sqrt{\frac{3}{2}}\frac{1}{q}\ $ &\ $\pm
\sqrt{-1+\frac{1}{q^2}}$ &\quad $-1\leq q\leq 1,\, q\neq 0$ \quad
& \quad $
\begin{array}{cc}
 -\frac{1}{2}-\frac{1}{2}\mu_0 & -\frac{1}{2}+\frac{1}{2}\mu_0 \\
  \left\{\pm\mu_1,1\right\} & \left\{\pm\mu_2,1\right\}
\end{array}
$&\quad unstable&\quad $\frac{3}{q^2}-1$&\quad0&\quad arbitrary\quad \vspace{0.2cm}\\
\hline $P_{7,8}$& 0 &\ $\pm i$ &\quad always & \quad $
\begin{array}{cc}
 4 & -1 \\
 \left\{\pm\frac{2}{5} i \sqrt{6} q,1\right\} & \{1,0\}
\end{array}
$&\quad unstable&\quad arbitrary &\quad 1&\quad $1/3$\quad \vspace{0.2cm}\\
\hline \hline
\end{tabular}
\end{center}
\caption[crit]{\label{crit2} The critical points of  a non-flat
universe with $\Lambda=0$ (case 2) and their behavior. We use the
notations $\mu_0=\sqrt{-15+\frac{16}{q^2}}$, $\mu_1=\frac{-9
q^2-\sqrt{16 q^2-15 q^4}+8}{4
   \sqrt{\frac{6}{q^2}-6} q}$ and $\mu_2=\frac{-9 q^2+\sqrt{16
   q^2-15 q^4}+8}{4 \sqrt{\frac{6}{q^2}-6} q}.$}
\end{table*}

In order to present the aforementioned behavior more
transparently, we evolve the autonomous system (\ref{eqzcase1})
numerically for  the choice $q=0.6$, and the results are shown in
figure \ref{Fig1}. As we can wee, in this case the critical point
$P_3$ is the global attractor of the system.

\subsection{Case 2: non-flat universe with $\Lambda=0$}\label{section B}

Under this scenario, and using the auxiliary variables
(\ref{auxilliaryx}),(\ref{auxilliaryy}),(\ref{auxilliaryz}),(\ref{auxilliaryu}),
the Friedmann equations (\ref{Fr1}), (\ref{Fr2}) become:
\begin{eqnarray}
&&1=x^2+y^2-z^2 \label{Fr1bcase2}
\\
&&\frac{H'}{H}=-3 x^2+2 z^2, \label{Fr2bcase2}
\end{eqnarray}
while the autonomous system writes:
\begin{eqnarray}
x'&=&x \left(3 x^2-2 z^2-3\right)+\sqrt{6} q \left(-x^2+z^2+1\right),\label{eqxcase2}\nonumber\\
z'&=& z \left[3 x^2-2 \left(z^2+1\right)\right].\label{eqzcase2}
\end{eqnarray}
It is defined in the phase space $\Psi=\left\{(x,z): x^2-z^2\leq
1, z\in\mathbb{R}\right\}$ and as before the phase space is not
compact. Finally, the matrix ${\bf {Q}}$ of the linearized
perturbation equations is:
\[{\bf {Q}}= \left[ \begin{array}{cc}
 9 x^2-2 \sqrt{6} q x-2 z^2-3\ \ \ \  & 2 \left(\sqrt{6} q-2 x\right) z \\
 6 x z & 3 x^2-6 z^2-2
\end{array} \right].\] The critical points, the eigensystems, the conditions for
their existence and stability, and the physical quantities are
presented in  table \ref{crit2}. Thus, $P_{1,2,3}$ are exactly the
same as in case 1, while $P_{5,6}$ are  saddle points except if
$q^2\rightarrow 1$, where they are nonhyperbolic. It is
interesting to notice that this scenario admits two more unstable
critical points, namely $P_{7,8}$, in which $z_c^2=-1$. These
points are of great physical importance, as we are going to see in
the next section.

\begin{figure}[ht]
\begin{center}
\mbox{\epsfig{figure=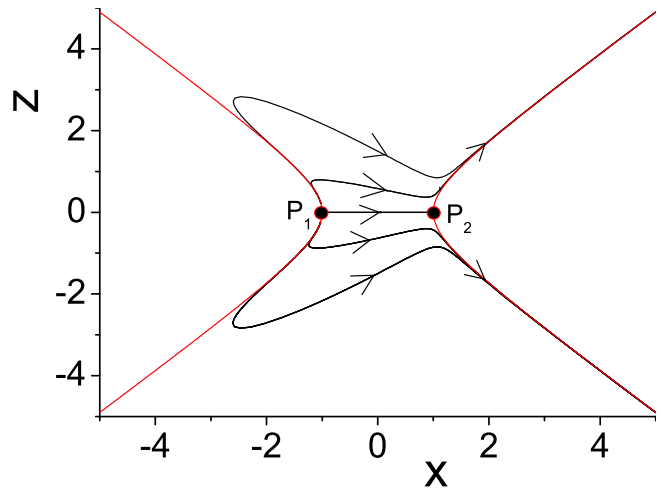,width=8.8cm,angle=0}} \caption{
(Color Online){\it Phase plane for a non-flat universe with
$\Lambda=0$ (case 2), for the choice $q=\sqrt{3}$. In this
specific scenario the critical points $P_3$ and $P_{5,6,7,8}$ do
not exist, while
 $P_{1}$ and $P_2$ are unstable (source and saddle respectively).}} \label{Fig2a}
\end{center}
\end{figure}

\begin{figure}[ht]
\begin{center}
\mbox{\epsfig{figure=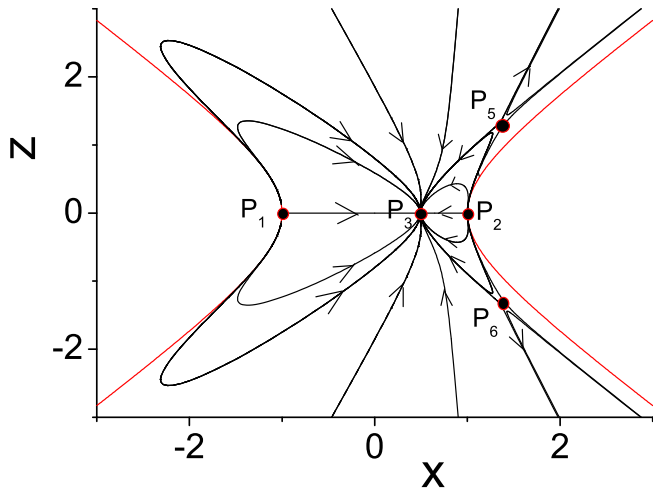,width=8.8cm,angle=0}} \caption{
(Color Online){\it Phase plane for a non-flat universe with
$\Lambda=0$ (case 2), for the choice $q=0.6$. In this specific
scenario the critical point $P_3$ is a local attractor, while
$P_{1,2}$ are unstable (sources) and $P_{5,6}$ are saddle ones. }}
\label{Fig2b}
\end{center}
\end{figure}

In order to present the results more transparently, in fig.
\ref{Fig2a} we present the numerical evolution of the system for
the choice $q=\sqrt{3}$. In this specific realization of the
scenario the critical points $P_3$ and $P_{5,6,7,8}$ do not exist.
We find only the source $P_{1}$ and the saddle $P_2$, and we
indeed observe that there is one orbit  approaching $P_2$ (the
solution with $z\equiv 0$). Finally, note that  the divergence of
the orbits towards the future is typical and suggests that the
future attractor of the system can be located at infinite regions.
In fig. \ref{Fig2b} we depict the phase-space graph for the choice
$q=0.6$. In this case the critical points $P_{1,2}$ are unstable
(sources), while $P_3$ is a local attractor. The points $P_{5,6}$
are saddle ones, and thus we observe that some orbits coming from
infinity spend a large amount of time near them before diverge
again in a finite time.

\subsection{Case 3: flat universe with $\Lambda \neq 0$}

In this case the Friedmann equations (\ref{Fr1}), (\ref{Fr2})
write as
\begin{eqnarray}
&&1=x^2+y^2-u^2 \label{Fr1bcase3}
\\
&&\frac{H'}{H}=-3 x^2, \label{Fr2bcase3}
\end{eqnarray}
and the autonomous system becomes:
\begin{eqnarray}
x'&=&\sqrt{6} q \left(u^2-x^2+1\right)+3 x \left(x^2-1\right),\label{eqxcase3}\nonumber\\
u'&=& 3 u x^2,\label{equcase3}
\end{eqnarray}
 defined
in the phase space $\Psi=\left\{(x,u): x^2-u^2\leq 1,
u\in\mathbb{R}\right\}$. As before the phase space is not compact.
The matrix ${\bf {Q}}$ of the linearized perturbation equations
is:
\[{\bf {Q}}= \left[ \begin{array}{cc}
 9 x^2-2 \sqrt{6} q x-3 \ \ &\ 2 \sqrt{6} q u \\
 6 u x & 3 x^2
\end{array} \right].\]
 The critical points,  the eigensystems, the conditions for
their existence and stability, and the physical quantities are
presented in  table \ref{crit3}.
\begin{table*}[t]
\begin{center}
\begin{tabular}{ccccccccc}
\hline \hline
 Cr. P &$x_c$&$u_c$&\quad Existence\quad& Eigensystem&\quad Stable for&\quad $w_M$&\quad $\Omega_{DE}$&\quad $w_{DE}$\\
\hline $P_{9,10}$& $\pm 1$ &0&\quad All $q$ \quad & \quad $
\begin{array}{cc}
  \mp 2 \sqrt{6} q+6 & 3\\
  \{1,0\} & \{0,1\}
\end{array}
$\quad& unstable&\quad1&\quad0&\quad arbitrary\quad \vspace{0.2cm}\\
\hline $P_{11}$& $\sqrt{\frac{2}{3}}\,q$ &0&\quad
$-\sqrt{\frac{3}{2}}<q<\sqrt{\frac{3}{2}}$ \quad & \quad $
\begin{array}{cc}
  -3+2 q^2 & 2q^2\\
  \{1,0\} & \{0,1\}
\end{array}
$& unstable &\quad$\frac{4}{3}q^2-1$&\quad0&arbitrary\quad \\
\vspace{-0.35cm}\\
\hline $P_{12,13}$& 0 &\ $\pm i$ &\quad always & \quad $
\begin{array}{cc}
 -3 & 0 \\
 \{1,0\} & \left\{2 i \sqrt{\frac{2}{3}} q,1\right\}
\end{array}
$&\quad NH&\quad arbitrary &\quad 1&\quad $-1$\quad \vspace{0.2cm}\\
\hline \hline
\end{tabular}
\end{center}
\caption[crit]{\label{crit3} The critical points of a flat
universe with $\Lambda\neq0$ (case 3) and their behavior. NH
stands for nonhyperbolic.}
\end{table*}
Note that the critical point $P_{11}$ is nonhyperbolic if
$q^2\in\{0,\,3/2\}$, while it is a saddle otherwise, with stable
(unstable) manifold tangent to the x- (u-) axis. Finally, the
system admits two more nonhyperbolic critical points, namely
$P_{12,13}$, in which $u_c^2=-1$.
\begin{figure}[ht]
\begin{center}
\mbox{\epsfig{figure=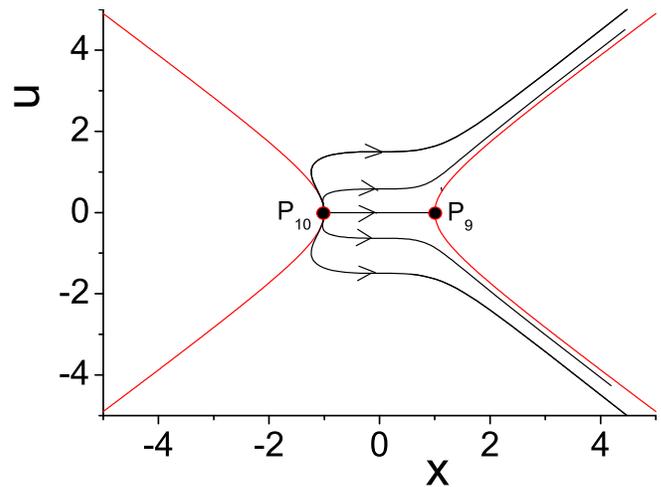,width=8.8cm,angle=0}} \caption{
(Color Online){\it Phase plane for a flat universe with
$\Lambda\neq0$ (case 3), for the choice $q=\sqrt{3}$. In this
specific scenario the critical point $P_{11}$ does not exists.
$P_{10}$ is unstable (source), while $P_9$ is a saddle one. }}
\label{Fig3a}
\end{center}
\end{figure}

In fig. \ref{Fig3a} we present the phase-space graph of the system
for the choice $q=\sqrt{3}$. In this case the critical point
$P_{11}$ does not exists, while $P_9$ and $P_{10}$ are unstable
(source and saddle respectively). The divergence of the orbits
towards the future is typical and suggests that the future
attractor of the system will be located at infinite regions.
\begin{figure}[ht]
\begin{center}
\mbox{\epsfig{figure=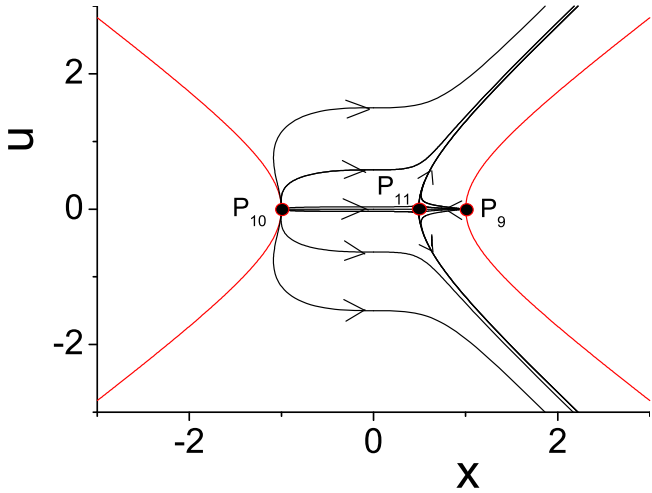,width=8.8cm,angle=0}} \caption{
(Color Online){\it  Phase plane for a flat universe with
$\Lambda\neq0$ (case 3), for the choice $q=0.6$. In this specific
scenario the critical point $P_{11}$ is a saddle one, while $P_9$
and $P_{10}$ are unstable (sources). }} \label{Fig3b}
\end{center}
\end{figure}
Finally, in fig. \ref{Fig3b} we display the phase-space graph for
the choice $q=0.6$. In this case the critical point $P_{11}$ is a
saddle one (with stable manifold tangent to the x-axis), while
$P_9$ and $P_{10}$ are unstable (sources). There are two orbits,
one joining $P_{10}$ with $P_{11}$ and one joining $P_9$ with
$P_{11},$ both of them overlapping the $x$-axis. Note that some
orbits remain close to $P_{11}$ before finally diverge towards the
future, and this suggests that the future attractor of the system
is located at infinite regions.

\subsection{Case 4: $k\neq 0, \Lambda \neq 0$}

\begin{table*}[t]
\begin{center}
\begin{tabular}{ccccccccc}
\hline \hline
 Cr. P &$x_c$& $z_c$&$u_c$&\quad Existence\quad&\quad Stable for&\quad $w_M$&\quad $\Omega_{DE}$&\quad $w_{DE}$\\
\hline $P_{14,15}$& $\pm 1$ &0&0&\quad All $q$  & \quad unstable & \quad1&\quad0&\quad arbitrary\quad \vspace{0.2cm}\\
\hline $P_{16}$& $\sqrt{\frac{2}{3}}\,q$ &0&0&\quad
$-\sqrt{\frac{3}{2}}<q<\sqrt{\frac{3}{2}}$  & \quad unstable &\quad $\frac{4}{3}q^2-1$&\quad0&arbitrary\quad \\
\vspace{-0.35cm}\\
\hline $P_{17,18}$& $\sqrt{\frac{3}{2}}\frac{1}{q}$ & $\quad\pm
\sqrt{-1+\frac{1}{q^2}}$ & $0$  &\quad $-1\leq q\leq 1,\, q\neq 0$  & \quad unstable & 2 &\quad $1-\frac{1}{q^2}$&\quad $1/3$\quad \vspace{0.2cm}\\
\hline $P_{19,20}$& 0 & 0 & $\pm i$  & always & \quad nonhyperbolic &\quad arbitrary &\quad 1  & \quad $-1$ \vspace{0.2cm}\\
\hline $P_{21,22}$& 0 & $\pm i$ & 0  & always & \quad unstable & \quad arbitrary &\quad 1  & \quad $1/3$ \vspace{0.2cm}\\
\hline \hline
\end{tabular}
\end{center}
\caption[crit]{\label{crit4} The  critical points of a non-flat
universe with $\Lambda\neq0$ (case 4) and their behavior.}
\end{table*}
Under this scenario, and using the auxiliary variables
(\ref{auxilliaryx}),(\ref{auxilliaryy}),(\ref{auxilliaryz}),(\ref{auxilliaryu}),
the Friedmann equations (\ref{Fr1}), (\ref{Fr2}) become:
\begin{eqnarray}
&&1=x^2+y^2-(u-k z)^2 \label{Fr1bcase4}
\\
&&\frac{H'}{H}=-3 x^2 + 2 z (-u + z). \label{Fr2bcase4}
\end{eqnarray}
while the autonomous system writes:
\begin{eqnarray}
x'&=&\sqrt{6} q \left[-x^2+(u-z)^2+1\right]+x \left[3 x^2+2 (u-z) z-3\right],\label{eqxcase4}\nonumber\\
z'&=&z \left[3 x^2+2 (u-z) z-2\right],\label{eqzcase4}\nonumber\\
u'&=& u \left[3 x^2+2 (u-z) z\right],\label{equcase4}
\end{eqnarray} defined
in the phase space $\Psi=\left\{(x,z,u): x^2-(u-k z)^2\leq 1, u,
z\in\mathbb{R}\right\}$, which is not compact. The linearized
perturbation matrix ${\bf {Q}}$ reads:
\begin{widetext}
\[{\bf {Q}}= \left[ \begin{array}{ccc}
 9 x^2-2 \sqrt{6} q x+2 (u-z) z-3\  &\  -2 \sqrt{6} k q u+2 x u+2 \sqrt{6} q z-4 x z\  &\ 2
   \left[x z+\sqrt{6} q (u-k z)\right] \\
 6 x z & 3 x^2-6 z^2+4 u z-2 & 2 z^2 \\
 6 u x & 2 u (u-2 z) & 3 x^2-2 z^2+4 u z
\end{array} \right].\]
\end{widetext}
 The critical points, and their corresponding information, are
presented in  table \ref{crit4}.

The critical point $P_{14}$ is nonhyperbolic if $q= \sqrt{3/2},$
it is a source if $q< \sqrt{3/2}$ or a saddle otherwise, while
$P_{15}$ is nonhyperbolic $q= -\sqrt{3/2},$ it is a source if $q>
-\sqrt{3/2}$ or a saddle otherwise. $P_{16}$ is nonhyperbolic if
$q^2\in\{0, \, 1,\ 3/2\}$ and saddle otherwise, while $P_{17,18}$
are nonhyperbolic if $q^2\rightarrow 1,$ and saddle otherwise. The
points $P_{19,20}$ have the eigenvalues $\{-3, -2, 0\}$ with
associated eigenvectors $\{1,0,0\}, \{0,1,1\}, \left\{\pm 2 i
\sqrt{\frac{2}{3}} q,0,1\right\}$. Hence, they are nonhyperbolic
possessing a 2-dimensional stable manifold. Finally, $P_{21,22}$
are unstable because they give rise to the eigenvalues
$\{4,2,-1\}$, with associated eigenvectors $\left\{-\frac{2}{5} i
\sqrt{6} q,1,0\right\}, \{0,1,1\}, \{1,0,0\}$.

\section{Beyond detailed balance: phase space
analysis}\label{nondetbal}

In this section we extend the phase-space analysis to a universe
governed by Ho\v{r}ava gravity in which the detailed balance
condition has been relaxed. In order to transform the
corresponding cosmological equations into an autonomous dynamical
system, we use the auxiliary variables $x$ and $y$ defined in
(\ref{auxilliaryx}),(\ref{auxilliaryy}), and furthermore we define
the following four new ones:
\begin{eqnarray}
&&x_1=\frac{\sigma_1}{3 (3 \lambda -1) H^2},\nonumber\\
&&x_2=\frac{k \sigma_2}{3 (3 \lambda -1) a^2 H^2},\nonumber\\
&&x_3=\frac{\sigma_3}{3 (3 \lambda -1) a^4 H^2}\nonumber\\
&&x_4=\frac{2 k \sigma_4}{(3 \lambda -1) a^6 H^2}.
 \label{auxilliaryc}
\end{eqnarray}
Thus, using these variables and the definitions (\ref{rhoDEext})
and (\ref{pDEext}), we can express the dark energy density and
equation-of-state parameters respectively  as:
\begin{eqnarray}
&& \Omega_{DE}|_{_\text{non-db}}\equiv\frac{2}{(3\la-1)H^2}\left(
\frac{\sigma_1}{6}+\frac{\sigma_3k^2}{6a^4} +\frac{\sigma_4k}{a^6}
 \right)=\nonumber\\
 &&\ \ \ \ \ \ \ \ \ \ \ \ \ \ \ =x_1 + x_3 + x_4,
  \label{OmegaDEext}
 \end{eqnarray}
\begin{equation}
  w_{DE}|_{_\text{non-db}}\equiv \frac{-\frac{\sigma_1}{6}+\frac{\sigma_3k^2}{18a^4}
+\frac{\sigma_4k}{6a^6} }{
\frac{\sigma_1}{6}+\frac{\sigma_3k^2}{6a^4}
+\frac{\sigma_4k}{a^6}}=-\frac{6 x_1 - 2 x_3 - x_4}{6 (x_1 - x_3 +
x_4)}
 \label{wDEext}.
\end{equation}
Note that the corresponding quantities for dark matter coincide
with those of the detailed balance case (expressions
(\ref{OmegaM}) and (\ref{wM})).

  Using the aforementioned auxiliary variables,
   the Friedmann equations (\ref{Fr1c}), (\ref{Fr2c}) become:
\begin{eqnarray}
&&1=x_1 + x_2 + x_3 + x_4 + x^2 + y^2\label{Fr1bextendedcase}
\\
&&\frac{H'}{H}=-3 x^2 - x_2 - 2 x_3 - 3 x_4.
\label{Fr2bextendedcase}
\end{eqnarray}
Thus, after using the first of these relations in order to
eliminate one variable, the corresponding autonomous system
writes:
\begin{eqnarray}
&&x_2'=2 x_2 \left(3 x^2+x_2+2 x_3+3
x_4-1\right),\label{ext2}\nonumber\\
&&x_3'=2x_3 \left(3 x^2+x_2+2 x_3+3 x_4-2\right),\label{ext3}\nonumber\\
&&x_4'=2 x_4 \left(3x^2+x_2+2 x_3+3
   x_4-3\right),\label{ext4}\nonumber\\
&& x'=   3 x^3+(x_2+2 x_3+3 x_4-3) x+\sqrt{6}
q y^2,\label{extx}\nonumber\\
&& y'=\left(3 x^2-\sqrt{6}
   q x+x_2+2 x_3+3 x_4\right) y,\label{exty}
\end{eqnarray}
defining a dynamical system in $\mathbb{R}^5$. Its critical points
and their properties are displayed in table \ref{critext}  and in
table \ref{densities} we present the corresponding observable
cosmological quantities.
\begin{table*}[t]
\begin{center}
\begin{tabular}{ccccccccc}
\hline \hline
 Cr. P &${x_2}_c$&${x_3}_c$&${x_4}_c$&${x}_c$&${y}_c$&\quad Existence\quad& Eigenvalues&\quad Stable for\\
\hline $P_{23}$& 0 & 0 & $1-{{x}_c}^2$ & ${x}_c$ & 0 & All $q$ &
$
\begin{array}{ccccc}
 6 & 0 & 2 & 4 & 3-\sqrt{6} q x_c
\end{array}
$ & NH\vspace{0.2cm}\\
\hline $P_{24,25}$& 0 & 0 & $0$ & $\pm 1$ & 0 & All $q$ & $
\begin{array}{ccccc}
 6 & 4 & 2 & 0 & 3\mp\sqrt{6} q
\end{array}
$ & NH\vspace{0.2cm}\\
\hline $P_{26}$& 0 & 0 & $0$ & 0 & 0 & All $q$ & $
\begin{array}{ccccc}
 -6 & -4 & -3 & -2 & 0
\end{array}
$ & stable\vspace{0.2cm}\\
\hline $P_{27,28}$& 0 & 0 & 0 & $\sqrt{\frac{2}{3}} q$ &
$\pm\sqrt{1-\frac{2 q^2}{3}}$ &  $q^2\leq \frac{3}{2}$ & $
\begin{array}{ccccc}
 4 q^2 & -3+2 q^2 & 2(-3+2 q^2) & 4(-1+q^2) & 2(-1+2 q^2)
\end{array}
$ & unstable\vspace{0.2cm}\\
\hline $P_{29}$& 1 & 0 & 0 & 0 & 0 & All $q$ & $
\begin{array}{ccccc}
 -4 & -2 & -2 & 2 & 1
\end{array}
$ & unstable\vspace{0.2cm}\\
\hline $P_{30,31}$& $1-\frac{1}{2q^2}$ & 0 & 0 &
$\frac{1}{\sqrt{6}q}$ & $\pm\frac{1}{\sqrt{3}q}$ & $q\neq 0$ & $
\begin{array}{ccccc}
 -4 & -2 & 2 & -1-\sqrt{-3+\frac{2}{q^2}} &
 -1+\sqrt{-3+\frac{2}{q^2}}
\end{array}
$ & unstable\vspace{0.2cm}\\
\hline $P_{32}$& 0 & 1 & 0 & 0 & 0 & All $q$ & $
\begin{array}{ccccc}
 -4 & -2 & 2 & 2 &
 -1
\end{array}
$ & unstable\vspace{0.2cm}\\
\hline $P_{33,34}$& 0 & $1-\frac{1}{q^2}$ & 0 &
$\frac{\sqrt{\frac{3}{2}}}{q}$ & $\pm\frac{1}{\sqrt{3}q}$ &
$q\neq 0$ & $
\begin{array}{ccccc}
 4 & -2 & 2 & -\frac{1}{2}\left(1-\sqrt{-15+\frac{16}{q^2}}\right) &
 -\frac{1}{2}\left(1+\sqrt{-15+\frac{16}{q^2}}\right)
\end{array}
$ & unstable\vspace{0.2cm}\\
\hline $P_{35}$& 0 & 0 & $1-\frac{3}{2 q^2}$ &
$\frac{\sqrt{\frac{3}{2}}}{q}$ & 0 & All $q$ & $
\begin{array}{ccccc}
 6 & 0 & 0 & 2 &
 4
\end{array}
$ & NH\vspace{0.2cm}\\
\vspace{-0.35cm}\\
\hline \hline
\end{tabular}
\end{center}
\caption[crit]{\label{critext} The critical points of a universe
governed by Ho\v{r}ava gravity beyond detailed balance (system
\ref{exty})) and their behavior. NH stands for nonhyperbolic.}
\end{table*}

\begin{table}[t]
\begin{center}
\begin{tabular}{ccccccccc}
\hline \hline
 Cr. P &\quad $w_M$&\quad $\Omega_M$\quad& \quad$\Omega_{DE}$&\quad $w_{DE}$\\
\hline $P_{23}$ & 1 & ${x_c}^2$ & $1 - {x_c}^2$ & $1/6$ \vspace{0.2cm}\\
\hline $P_{24,25}$ &1 & 1 & 0 & arbitrary \vspace{0.2cm}\\
\hline $P_{26}$ & arbitrary& 0& 1& -1 \vspace{0.2cm}\\
\hline $P_{27,28}$ & $\frac{4 q^2}{3}-1$ &1 &$\frac{-2
q^2+\sqrt{9-6 q^2}+3}{-2
q^2+\sqrt{9-6 q^2}+6}$ &-1 \vspace{0.2cm}\\
\hline $P_{29}$ & 1 &1  &0 & arbitrary \vspace{0.2cm}\\
\hline $P_{30,31}$ & $-\frac{1}{3}$ & $\frac{1}{2 q^2}$ &
$\frac{\sqrt{3} q+1}{3 q^2+\sqrt{3} q+1}$ & -1 \vspace{0.2cm}\\
\hline $P_{32}$ & arbitrary &0 & 1& 1/3 \vspace{0.2cm}\\
\hline $P_{33,34}$& $\frac{1}{3}$ & $\frac{1}{q^2}$ &
$1+\frac{3}{\left(\sqrt{3}-3 q\right) q-1}$ &$\frac{2-q
   \left(q+\sqrt{3}\right)}{\left(\sqrt{3}-3 q\right) q+2}$\vspace{0.2cm}\\
\hline $P_{35}$ &1 & $\frac{3}{2 \text{q$\phi $}^2}$ & $1-\frac{3}{2 \text{q$\phi $}^2}$ &$\frac{1}{6}$\vspace{0.2cm}\\
\vspace{-0.35cm}\\
\hline \hline
\end{tabular}
\end{center}
\caption[crit]{\label{densities} Observable cosmological
quantities of a universe governed by Ho\v{r}ava gravity beyond
detailed balance.}
\end{table}

The curve of nonhyperbolic critical points denoted by $P_{23}$ is
``normally hyperbolic'' \cite{normally}. Thus, examining the sign
of the real parts of the non-null eigenvalues, we find that they
are always local sources provided $q x_c>\sqrt{6}/2$.
 $P_{27,28}$ are saddle points
and their stable manifold can be 4-dimensional provided
$-\frac{\sqrt{2}}{2}<q<\frac{\sqrt{2}}{2}$. $P_{29}$ has a
2-dimensional unstable manifold tangent to the $x_2$-$y$ plane and
its stable manifold is always 3-dimensional. $P_{30,31}$ have a
4-dimensional stable manifold provided $q^2>\frac{2}{3}$ or
$-\sqrt{\frac{2}{3}}\leq q\leq -\frac{\sqrt{2}}{2}$ or
$\frac{\sqrt{2}}{2}< q\leq \sqrt{\frac{2}{3}}$. Finally,
$P_{33,34}$ has a 3-dimensional stable manifold if
$q^2>\frac{16}{15}$ or $-{\frac{4}{\sqrt{15}}}\leq q< -1$ or $1<
q\leq {\frac{4}{\sqrt{15}}}$.

Amongst all these critical points $P_{26}$ although nonhyperbolic,
proves to be a stable one. To see that  we use techniques such as
the Normal Forms theorem \cite{arrowsmith,wiggins}, which allows
to obtain a simplified system by successive nonlinear
transformations. In particular, we first  make the linear
transformation $(x_2,x_3,x_4,x,y)\rightarrow (x, x_3, x_2, x_4,
y)$ in order to transform the matrix of linear perturbations $Q$
evaluated in $(0,0,0,0,0)$ to its Jordan real form:
$\text{diag}\left(-6,-4,-3,-2,0\right).$ The next step is to
perform a quadratic coordinate transformation given by
$$\left(
\begin{array}{c}
 x_2\\
 x_3\\
 x_4\\
  x\\
 y
\end{array}
\right)\rightarrow \left(
\begin{array}{c}
x_2 -x_2 (x+x_2+x_3)\\
x_3 -x_3
   (x+x_2+x_3)\\
x_4   \sqrt{\frac{2}{3}} q
   y^2-\frac{1}{2} (x+x_2+x_3) x_4\\
x   -x
   (x+x_2+x_3)\\
y   -\frac{1}{6} \left(3 x+3 x_2+3
   x_3-2 \sqrt{6} q x_4\right) y
\end{array}
\right),$$ which eliminates the second order terms (all of which
are non-resonant). Finally, we implement the cubic transformation
\begin{widetext}
$$\left(
\begin{array}{c}
 x_2\\
 x_3\\
 x_4\\
  x\\
 y
\end{array}
\right)\rightarrow \left(
\begin{array}{c}
x_2+x_2
   \left[(x+x_2+x_3)^2-x_4^2\right]\\
x_3+   x_3
   \left[(x+x_2+x_3)^2-x_4^2\right]\\
x_4+   \frac{1}{24}
   \left[-12 x_4^3+9 (x+x_2+x_3)^2 x_4+96
   q^2 y^2 x_4+8 \sqrt{6} q
   (x_3-2 x) y^2\right]\\
x+   x
   \left[(x+x_2+x_3)^2-x_4^2\right]\\
y+   \frac{y \left\{945
   x^2+42 \left(45 x_2+45 x_3-16 \sqrt{6} q
   x_4\right) x+5 \left\{189 x_2^2+14 \left(27 x_3-8
   \sqrt{6} q x_4\right) x_2+3 \left\{63
   x_3^2-40 \sqrt{6} q x_4 x_3+28
   \left[\left(2 q^2-3\right) x_4^2-12 \text{q$\phi
   $}^2 y^2\right]\right\}\right\}\right\}}{2520}
\end{array}
\right).$$ \end{widetext}
 As a result, we get the simplified system in the new
variables:
\begin{eqnarray}
&&x_2'=-6 x_2+\mathcal{O}{(4)},\label{normal2}\nonumber\\
&&x_3'=-4x_3 +\mathcal{O}{(4)},\label{normal3}\nonumber\\
&&x_4'=x_4 \left(-3+4 q^2
y^2\right)+\mathcal{O}{(4)},\label{normal4}\nonumber
\end{eqnarray}
\begin{eqnarray}
&& x'=-2x+\mathcal{O}{(4)},\label{normalx}\nonumber\\
&& y'=-2 q^2 y^3+\mathcal{O}{(4)},\label{normaly}
\end{eqnarray}
where $\mathcal{O}{(4)}$ denotes terms of fourth order with
respect to the vector norm. The center manifold, $W^c_\text{loc},$
is tangent to the y-axis at the origin, and it can be represented
locally, up to an error $\mathcal{O}{(4)}$, as the graph
\begin{eqnarray}
&&W^c_\text{loc}=\left\{(x_2,x_3,x_4,x,y)\in\mathbb{R}^5:
x_2=x_{20}e^{-\frac{3}{2 q^2 y^2}},\right.\nonumber\\
&&\left.\ \ \ \ \ \ \ \ \ \ \ \ \ \  x_3=x_{30}e^{-\frac{1}{q^2
y^2}},x_4=x_{40}{y^{-2}}e^{-\frac{1}{q^2
y^2}},\right.\nonumber\\
&&\left.\ \ \ \ \ \ \ \ \ \, \ \ \ \ \ \ x=x_{0}e^{-\frac{1}{2 q^2
y^2}}, |y|<\varepsilon\right\},
\end{eqnarray}
where $\varepsilon$ is a positive, sufficiently small constant.
From (\ref{normaly}) we deduce that $y$, under the initial
condition $y(0)=y_0$, evolves  as $y(M)=y_0 (1+4 q^2 y_0^2
M)^{-1/2}$. Hence, as time passes the origin is approached and
thus the critical point $P_{26}$ is definitely an attractor. Note
also that since it has a 1-dimensional center manifold tangent to
the $y$-axis, then the stable manifold of $P_{26}$ is
4-dimensional, which also proves that $P_{26}$ is a late-time
attractor.

\section{Cosmological implications}
\label{cosmimpl}

Since we have performed a phase-space analysis of a universe
governed by Ho\v{r}ava-Lifshitz gravity, with or without the
detailed-balance condition, we can now discuss the corresponding
cosmological behavior.

\subsection{Detailed balance}\label{section O2}

\subsubsection{Case 1: flat universe with
$\Lambda=0$}\label{section A2}

In this scenario the critical points $P_{1,2}$ are not relevant
from a cosmological point of view, since apart from being unstable
they correspond to complete dark matter domination, with the
matter equation-of-state parameter being unphysically stiff.
However, point $P_3$ is more interesting since it is stable for
$-1<q<1$ and thus it can be the late-time state of the universe.
If additionally we desire to keep the dark-matter
equation-of-state parameter in the physical range $0<w_M<1$ then
we have to restrict the parameter $q$ in the range
$\sqrt{3}/2<q<\sqrt{3/2}$. However, even in this case the universe
is finally completely dominated by dark matter. The fact that
$z_c=0$ means that in general this sub-class of universes will be
expand forever. The critical points $P_4$ consist a stable
late-time solution, with a physical dark-matter equation-of-state
parameter $w_M=1/3$, but with zero dark energy density. We mention
that the dark-matter domination of the case at hand was expected,
since in the absent of curvature and of a cosmological constant
the corresponding Ho\v{r}ava-Lifshitz universe is comprised only
by dark matter. Note however that the dark-energy
equation-of-state parameter can be arbitrary.

\subsubsection{Case 2: non-flat universe with $\Lambda=0$}\label{section B2}

In this scenario, the first three critical points are identical
with those of case 1, and thus the  physical implications are the
same. The critical points $P_{5,6}$ are unstable, corresponding to
a dark-matter dominated universe. This was expected since in the
absence of the cosmological constant $\Lambda$, the curvature role
is downgrading as the scale factor increases and thus in the end
this case tends to the case 1 above. Note however that at early
times, where the scale factor is small, the behavior of the system
will be significantly different than case 1, with the dark energy
playing an important role. This different behavior is observed in
the corresponding phase-space figures  \ref{Fig2a}, \ref{Fig2b}
comparing with figure \ref{Fig1}.

The case at hand admits another solution sub-class, namely points
$P_{7,8}$. In these points $z_c^2=-1$, and thus using
({\ref{auxilliaryz}}) we straightforwardly find the late-time
solution $a(t)=e^{i\gamma t}$, with
$\gamma=|\kappa^2\mu/[4(3\lambda-1)]|$. This solution corresponds
to an oscillatory universe \cite{cyclic.clifton,Saridakis:2007cf},
and in the context of Ho\v{r}ava-Lifshitz cosmology it has already
been studied in the literature
\cite{Brandenberger:2009yt,Brandenberger:2009ic,Cai:2009in}.
However, as we see, these critical points are unstable and thus
this solution subclass cannot be a late-time attractor in the case
of a non-flat universe with zero cosmological constant. This
situation will change in the case where the cosmological constant
is switched on.

\subsubsection{Case 3: flat universe with $\Lambda \neq 0$}\label{section C2}

Under this scenario, the Ho\v{r}ava-Lifshitz  universe admits two
unstable critical points ($P_{9,10}$), completely dominated by
stiff dark matter. Point $P_{11}$ exhibits a more physical dark
matter equation-of-state parameter, but still with negligible dark
energy at late times. The case at hand admits the two
nonhyperbolic points $P_{12,13}$ possessing $u_c^2=-1$, and thus
(as can be seen by ({\ref{auxilliaryu}})) they correspond to the
oscillatory solution $a(t)=e^{i\delta t}$, with
$\delta=|\kappa^2\mu\Lambda/[4(3\lambda-1)]|$. We mention that
these points are nonhyperbolic, with a negative eigenvalue, and
thus they have a large probability to be a late-time solution of
Ho\v{r}ava-Lifshitz universe. Additionally, they correspond to
dark-energy domination, with dark-energy equation-of-state
parameter $-1$ and an arbitrary $w_M$. These features make them
good candidates to be a realistic description of the universe. We
mention that this result is independent from the parameter $q$
which comes from the dark matter sector. Thus, we conclude that it
is valid independently of the matter-content of the universe.
Indeed, this behavior is novel, and arises purely by the extra
terms that are present in Ho\v{r}ava gravity.

\subsubsection{Case 4: non-flat universe with $\Lambda \neq 0$}\label{section D2}

This case admits the unstable critical points $P_{14,15,16}$ which
correspond to a dark-matter dominated universe, and the unstable
points $P_{17,18}$ which are unphysical since they possess
$w_M=2$. As expected, the system admits also the unstable points
$P_{21,22}$ which correspond to oscillatory universes with
$a(t)=e^{i\gamma t}$ ($\gamma=|\kappa^2\mu/[4(3\lambda-1)]|$).
However, we find two more oscillatory critical points, namely
$P_{19,20}$, which correspond to $a(t)=e^{i\delta t}$, with
$\delta=|\kappa^2\mu\Lambda/[4(3\lambda-1)]|$. These points are
nonhyperbolic, with a negative eigenvalue, and thus they have a
large probability to be the late-time state of the universe, and
additionally this result is independent of the specific form of
the dark-matter content. Furthermore, they correspond to a
dark-energy dominated universe, with $w_{DE}=-1$ and arbitrary
$w_M$. Thus, they are good candidates for a realistic description
of the universe.

\subsection{Beyond detailed balance}\label{section O2}

Let us now discuss about the cosmological behavior of a
Ho\v{r}ava-Lifshitz universe, in the case where the detailed
balance condition is abandoned. In this case the system admits the
unstable critical points $P_{27,28,29}$ which correspond to dark
matter domination, the unstable point $P_{32}$ corresponding to an
unphysical dark-energy dominated universe, and the unstable
$P_{30,31,33,34}$ which have physical $w_M$, $w_{DE}$ but
dependent on the specific dark-matter form. The system admits also
the critical points $P_{23}$, $P_{35}$ which are nonhyperbolic
with positive non-null eigenvalues, thus unstable, with
furthermore unphysical cosmological quantities. Additionally,
points $P_{24,25}$ are also dark-matter dominated, unstable
nonhyperbolic ones.

It is interesting to notice that since $\sigma_3$ has an arbitrary
sign, $P_{33,34}$ could also correspond to an oscillatory
universe, for a wide region of the parameters $\sigma_3$ and $q$.
However, this oscillatory behavior has a small probability to be
the late-time state of the universe because it is not stable (with
at least two positive eigenvalues). Additionally, the fact that it
depends on $q$ means that this solution depends on the matter form
of the universe.

The scenario at hand admits a final critical point, namely
$P_{26}$. As we showed in detail in section \ref{nondetbal} using
Normal Forms techniques, it is indeed stable and thus it can be a
late-time attractor of Ho\v{r}ava-Lifshitz universe beyond
detailed balance. Using the definition of the auxiliary variables,
we can straightforwardly show that it corresponds to an eternally
expanding solution. Additionally, it is characterized by complete
dark energy domination, with dark-energy equation-of-state
parameter $-1$ and arbitrary $w_M$. Note also that this result is
independent of the specific form of the dark-matter content. These
feature make it a very good candidate for the description of our
universe. We mention that according to the initial conditions,
this universe on its way towards this late-time attractor can be
just an expanding universe with a non-negligible dark matter
content, which is in agreement with observations, and this can be
verified also by numerical investigation. This fact makes the
aforementioned result more concrete.

\section{Conclusions}
\label{conclusions}

In this work we performed a detailed phase-space analysis of
Ho\v{r}ava-Lifshitz cosmology, with and without the
detailed-balance condition. In particular, we examined if a
universe governed by Ho\v{r}ava gravity can have late-time
solutions compatible with observations.

In the case where the detailed-balance condition is imposed, we
find that the universe can reach a bouncing-oscillatory state at
late times, in which dark-energy, behaving as a simple
cosmological constant, will be dominant. Such solutions were
already investigated in the context of Ho\v{r}ava-Lifshitz
cosmology
\cite{Brandenberger:2009yt,Brandenberger:2009ic,Cai:2009in} as
possible ones, but now we see that they can indeed be the
late-time attractor for the universe. They arise purely from the
novel terms of Ho\v{r}ava-Lifshitz cosmology, and in particular
the dark-radiation term proportional to $a^{-4}$ is responsible
for the bounce, while the cosmological constant term is
responsible for the turnaround.

In the case where the detailed-balance condition is abandoned, we
find that the universe reaches an eternally expanding solution at
late times, in which dark-energy, behaving like a cosmological
constant, dominates completely. Note that according to the initial
conditions, the universe on its way to this late-time attractor
can be an expanding one with non-negligible matter content. We
mention that this behavior is independent of the specific form of
the dark-matter content. Thus, the aforementioned features make
this scenario a good candidate for the description of our
universe, in consistency with observations. Finally, in this case
the universe has also a probability to reach an oscillatory
solution at late times, if the initial conditions lie in its basin
of attraction (in this case the eternally expanding solution will
not be reached).

Although this analysis indicates that Ho\v{r}ava-Lifshitz
cosmology can be compatible with observations, it does not
enlighten the discussion about possible conceptual and
phenomenological problems and instabilities of Ho\v{r}ava-Lifshitz
gravity,  nor it can interfere with the questions concerning the
validity of its theoretical background, which is the subject of
interest of other studies. It just faces the problem from the
cosmological point of view, and thus its results can been taken
into account only if  Ho\v{r}ava gravity passes successfully the
aforementioned theoretical tests.

\vskip .1in \noindent {\bf {Acknowledgments}}

 G. L wishes to thank
the MES of Cuba for partial financial support of this
investigation. His research was also supported by Programa
Nacional de Ciencias B\'asicas (PNCB).

\vskip .1in \noindent {\bf {Note added}}

While this work was being typed, we became aware of
\cite{Carloni:2009jc}, which presents also an analysis of the
phase-space of Ho\v{r}ava-Lifshitz cosmology, though in a
different framework. We agree with \cite{Carloni:2009jc} on the
regions of overlap.


\addcontentsline{toc}{section}{References}
\begin{thebibliography}{99}



\bibitem{hor2}
  P.~Horava,
  arXiv:0811.2217 [hep-th].

\bibitem{hor1}
  P.~Horava,
  JHEP {\bf 0903}, 020 (2009)
  [arXiv:0812.4287 [hep-th]].

\bibitem{hor3}
  P.~Horava,
  Phys.\ Rev.\  D {\bf 79}, 084008 (2009)
  [arXiv:0901.3775 [hep-th]].

\bibitem{hor4}
  P.~Ho\v rava,
  arXiv:0902.3657 [hep-th].

\bibitem{Volovik:2009av}
  G.~E.~Volovik,
  arXiv:0904.4113 [gr-qc].

\bibitem{Cai:2009ar}
  R.~G.~Cai, Y.~Liu and Y.~W.~Sun,
  arXiv:0904.4104 [hep-th].

\bibitem{Cai:2009dx}
  R.~G.~Cai, B.~Hu and H.~B.~Zhang,
  arXiv:0905.0255 [hep-th].

\bibitem{Orlando:2009en}
  D.~Orlando and S.~Reffert,
  arXiv:0905.0301 [hep-th].

\bibitem{Nishioka:2009iq}
  T.~Nishioka,
  arXiv:0905.0473 [hep-th].

\bibitem{Konoplya:2009ig}
  R.~A.~Konoplya,
  arXiv:0905.1523 [hep-th].

\bibitem{Charmousis:2009tc}
  C.~Charmousis, G.~Niz, A.~Padilla and P.~M.~Saffin,
  arXiv:0905.2579 [hep-th].

\bibitem{Sotiriou:2009bx}
  T.~P.~Sotiriou, M.~Visser and S.~Weinfurtner,
  arXiv:0905.2798 [hep-th].

\bibitem{Bogdanos:2009uj}
  C.~Bogdanos and E.~N.~Saridakis,
  arXiv:0907.1636 [hep-th].

\bibitem{Kluson:2009rk}
  J.~Kluson,
  arXiv:0907.3566 [hep-th].


\bibitem{Afshordi:2009tt}
  N.~Afshordi,
  arXiv:0907.5201 [hep-th].

\bibitem{Myung:2009ur}
  Y.~S.~Myung,
  arXiv:0907.5256 [hep-th].


\bibitem{Li:2009bg}
  M.~Li and Y.~Pang,
  arXiv:0905.2751 [hep-th].

\bibitem{Visser:2009fg}
  M.~Visser,
  arXiv:0902.0590 [hep-th].

\bibitem{Chen:2009bu}
  J.~Chen and Y.~Wang,
  arXiv:0905.2786 [gr-qc].

\bibitem{Chen:2009ka}
  B.~Chen and Q.~G.~Huang,
  arXiv:0904.4565 [hep-th].

\bibitem{Shu:2009gc}
  F.~W.~Shu and Y.~S.~Wu,
  arXiv:0906.1645 [hep-th].

\bibitem{Calcagni:2009ar}
  G.~Calcagni,
  arXiv:0904.0829 [hep-th].

\bibitem{Kiritsis:2009sh}
  E.~Kiritsis and G.~Kofinas,
  arXiv:0904.1334 [hep-th].

\bibitem{Lu:2009em}
  H.~Lu, J.~Mei and C.~N.~Pope,
  arXiv:0904.1595 [hep-th].

\bibitem{Nastase:2009nk}
  H.~Nastase,
  arXiv:0904.3604 [hep-th].

\bibitem{Colgain:2009fe}
  E.~O.~Colgain and H.~Yavartanoo,
  arXiv:0904.4357 [hep-th].

\bibitem{Ghodsi:2009rv}
  A.~Ghodsi,
  arXiv:0905.0836 [hep-th].

\bibitem{Minamitsuji:2009ii}
  M.~Minamitsuji,
  arXiv:0905.3892 [astro-ph.CO].

\bibitem{Ghodsi:2009zi}
  A.~Ghodsi and E.~Hatefi,
  arXiv:0906.1237 [hep-th].

\bibitem{Mukohyama:2009gg}
  S.~Mukohyama,
  arXiv:0904.2190 [hep-th].

\bibitem{Piao:2009ax}
  Y.~S.~Piao,
  arXiv:0904.4117 [hep-th].

\bibitem{Gao:2009bx}
  X.~Gao,
  arXiv:0904.4187 [hep-th].

\bibitem{Chen:2009jr}
  B.~Chen, S.~Pi and J.~Z.~Tang,
  arXiv:0905.2300 [hep-th].

\bibitem{Gao:2009ht}
  X.~Gao, Y.~Wang, R.~Brandenberger and A.~Riotto,
  arXiv:0905.3821 [hep-th].


\bibitem{Wang:2009yz}
  A.~Wang and R.~Maartens,
  arXiv:0907.1748 [hep-th].

\bibitem{Kobayashi:2009hh}
  T.~Kobayashi, Y.~Urakawa and M.~Yamaguchi,
  arXiv:0908.1005 [astro-ph.CO].



\bibitem{Mukohyama:2009zs}
  S.~Mukohyama, K.~Nakayama, F.~Takahashi and S.~Yokoyama,
  arXiv:0905.0055 [hep-th].

\bibitem{Takahashi:2009wc}
  T.~Takahashi and J.~Soda,
  arXiv:0904.0554 [hep-th].

\bibitem{Koh:2009cy}
  S.~Koh,
  arXiv:0907.0850 [hep-th].

\bibitem{Brandenberger:2009yt}
  R.~Brandenberger,
  arXiv:0904.2835 [hep-th].

\bibitem{Brandenberger:2009ic}
  R.~H.~Brandenberger,
  arXiv:0905.1514 [hep-th].

\bibitem{Cai:2009in}
  Y.~F.~Cai and E.~N.~Saridakis,
  arXiv:0906.1789 [hep-th].

\bibitem{Danielsson:2009gi}
  U.~H.~Danielsson and L.~Thorlacius,
  JHEP {\bf 0903}, 070 (2009)
  [arXiv:0812.5088 [hep-th]].

\bibitem{Cai:2009pe}
  R.~G.~Cai, L.~M.~Cao and N.~Ohta,
  arXiv:0904.3670 [hep-th].

\bibitem{Myung:2009dc}
  Y.~S.~Myung and Y.~W.~Kim,
  arXiv:0905.0179 [hep-th].

\bibitem{Kehagias:2009is}
  A.~Kehagias and K.~Sfetsos,
  arXiv:0905.0477 [hep-th];

\bibitem{Cai:2009qs}
  R.~G.~Cai, L.~M.~Cao and N.~Ohta,
  arXiv:0905.0751 [hep-th].

\bibitem{Mann:2009yx}
  R.~B.~Mann,
  arXiv:0905.1136 [hep-th].

\bibitem{Bertoldi:2009vn}
  G.~Bertoldi, B.~A.~Burrington and A.~Peet,
  arXiv:0905.3183 [hep-th].

\bibitem{Castillo:2009ci}
  A.~Castillo and A.~Larranaga,
  arXiv:0906.4380 [gr-qc].

\bibitem{BottaCantcheff:2009mp}
  M.~Botta-Cantcheff, N.~Grandi and M.~Sturla,
  arXiv:0906.0582 [hep-th].


\bibitem{Lee:2009rm}
  H.~W.~Lee, Y.~W.~Kim and Y.~S.~Myung,
  arXiv:0907.3568 [hep-th].



\bibitem{Saridakis:2009bv}
  E.~N.~Saridakis,
  arXiv:0905.3532 [hep-th].

\bibitem{Park:2009zr}
  M.~i.~Park,
  arXiv:0906.4275 [hep-th].

\bibitem{Wang:2009rw}
  A.~Wang and Y.~Wu,
  arXiv:0905.4117 [hep-th].

\bibitem{Appignani:2009dy}
  C.~Appignani, R.~Casadio and S.~Shankaranarayanan,
  arXiv:0907.3121 [hep-th].



\bibitem{Kim:2009dq}
  S.~S.~Kim, T.~Kim and Y.~Kim,
  arXiv:0907.3093 [hep-th].

\bibitem{Harko:2009qr}
  T.~Harko, Z.~Kovacs and F.~S.~N.~Lobo,
  arXiv:0908.2874 [gr-qc].


\bibitem{Copeland:1997et}
  E.~J.~Copeland, A.~R.~Liddle and D.~Wands,
  Phys.\ Rev.\  D {\bf 57}, 4686 (1998).

\bibitem{expon} P.G. Ferreira, M. Joyce, Phys. Rev. Lett.  {\bf79}, 4740
(1997).


\bibitem{expon1}
 E.J. Copeland, M. Sami, S. Tsujikawa, Int. J. Mod. Phys. D
 {\bf15}, 1753 (2006).

\bibitem{expon2}
  X.~m.~Chen, Y.~g.~Gong and E.~N.~Saridakis,
  JCAP {\bf 0904}, 001 (2009).




\bibitem{arrowsmith} D. K. Arrowsmith D. K. and Place C. M.,  {\it{An introduction to Dynamical Systems}}, Cambridge University Press (1990).

\bibitem{wiggins} S. Wiggins,  {\it{Introduction to Applied Nonlinear Dynamical Systems and Chaos -2nd Edition}}, Springer
(2003).


\bibitem{normally} B. Aulbach, {\it{Continuous and discrete dynamics near manifolds of
equilibria}}, Lecture Notes in Mathematics, No 1058,
Springer-Verlag (1981).

\bibitem{cyclic.clifton}
T.~Clifton and J.~D.~Barrow, Phys. Rev. D {\bf 75}, 043515 (2007)
[arXiv:gr-qc/0701070].


\bibitem{Saridakis:2007cf}
  E.~N.~Saridakis,
  Nucl.\ Phys.\  B {\bf 808}, 224 (2009)
  [arXiv:0710.5269 [hep-th]].




\bibitem{Carloni:2009jc}
  S.~Carloni, E.~Elizalde and P.~J.~Silva,
  arXiv:0909.2219 [hep-th].



\end{thebibliography}
\end{document}